# Digital hybridity and relics in cultural heritage: using corpus linguistics to inform design in emerging technologies from AI to VR

Emma McClaughlin,[1] Glenn McGarry,[2] Alan Chamberlain,[3] Geert De Wilde,[4] Oliver Butler.[5]

[1] School of English, University of Nottingham (emma.mcclaughlin@nottingham.ac.uk), Nottingham, UK. orcid.org/0000-0001-9659-2589

[2] School of Computer Science, University of Nottingham, Nottingham, UK. orcid.org/0000-0001-9518-3126

[3] School of Computer Science, University of Nottingham, Nottingham, UK. orcid.org/0000-0002-2122-8077

[4] Department of Modern Languages, Aberystwyth University, Aberystwyth, UK. orcid.org/0000-0003-4317-2956

[5] School of Law, University of Nottingham, Nottingham, UK. orcid.org/0009-0004-3979-6963

**Abstract**

Hybrid technologies enable the blending of physical and digital elements, creating new ways to experience and interact with the world. Such technologies can transform engagement with relics—both secular and sacred—but they present challenges for capturing faith, belief, and representation responsibly. Given the complexities of digital representation and the ethical challenges inherent in digitising culturally significant objects, a transdisciplinary understanding of these issues is needed. To inform this discussion from a linguistic perspective, we examined the representation of relics in historical and contemporary texts. Using a corpus linguistic approach to extract modifiers of the word 'relic' in corpora of Early Modern English books and contemporary web-sourced texts from 2021, we examined the multifaceted ways in which relics have been perceived and evaluated over time. Early texts consider relics as both objects of moral and spiritual significance, and tools of religious and political control, while they are more often framed as heritage symbols, reflecting past events, places, and traditions in contemporary texts. We discuss how hybrid, sometimes AI-based technologies can enhance accessibility and engagement, whilst also challenging traditional sensitivities around authenticity and sensory experience, which are integral to the meaning and significance of relics.

## 1. Background

The central question guiding our research is: how can humanities-based methods, such as corpus linguistics, inform the digitisation and subsequent interpretations of cultural heritage? This focus is critical for addressing issues of bias, historical interpretivism, and the role of context and truth in shaping authenticity and multiple interpretations of reality in this area. Further to this, our work aims to act as a catalyst for reflecting on the implications of using large language models (LLMs) in embodied, tangible technologies. Earlier research examined technology in musical instruments (Benford et al. 2023), which generated their own archives and documented diverse player interactions. This led us to explore 'relicing' - the deliberate aging of instruments to evoke vintage aesthetics despite their newness - and to explore the validity of the humanities as resource and approach to design in the context of AI (Woycicki et al. 2025). To further explore this previously identified gap in the research and explore this evolving researcher stance, this paper demonstrates the practical application of humanities-based approaches for establishing multiple historical perspectives and framings. Such approaches help designers to act responsibly to highlight different representations and meanings when creating artefacts, software, and exhibitions that embody layered meanings within a single physical artefact. This also has broader implications for AI developers, whose systems often present generalized, one-dimensional views rather than multifaceted notions of authenticity.

Corpus linguistics enables us to advance responsible research and innovation in the arts and humanities (Piskopani et al. 2023), moving beyond image-based approaches and deepening the concept of Explainability - an area of growing interest in AI and robotics. Our work supports interdisciplinary researchers at the intersection of AI, the arts, and explainability (Ford et al. 2025), encouraging new understandings grounded in linguistic analysis. This is particularly relevant for disciplines such as theology, linguistics, and digital humanities, where design practices must account for everyday values, beliefs, and reasoning, as well as the technologies entwined with them (Vitullo et al. 2025), and the aura artefacts can create (Radermacher 2019).

### 1.1. Digital technologies and cultural artefacts

The development of digital technologies has facilitated engagement with cultural artefacts in a digitally mediated way, blending online and offline environments and merging digital and physical experiences (Marini & Agostino, 2022, p. 606). Several studies have highlighted the role that virtual technologies can play in enriching cultural and natural heritage experiences. The EU Etruscanning project saw the Regolini Galssi tomb rendered in a permanent virtual reality (VR) installation in the Vatican Museums (see Pietroni, 2013, 2019). The installation was set up to enhance visitor experience after viewing the real objects from the tomb. In VR, the objects are shown in situ within the tomb and have been digitally restored in the virtual reconstruction. Similarly, 'Building Virtual Rome' was a virtual archaeology exhibition held at Trajan Markets in Rome, which aimed to bring the 'vision' of Ancient Rome to life (Pietroni, 2019, p. 12) by offering visitors the chance to experience artefacts through digital installations, videos, multimedia, and VR using stereoscopy, haptic devices, and artificial life technologies. The immersive Virtual Heritage app (Brundell et al., 2022), recreates a Roman forum at the site of Venta Icenorum in Norfolk, UK, aiming to deepen visitors' connections to the site by overlaying virtual reconstructions on the present-day landscape to enhance understanding and emotional engagement with the past. Finally, the VRtefacts project (Spence et al. 2020) used a performative substitutional reality approach to guide visitors to Derby Museum, UK through structured, narrative-driven interactions with artefacts. The performance-led mixed reality design of the study combined visitor interactions with 3D printed and virtually rendered objects. Such approaches address museum restrictions on handling delicate objects, digitally reinstate historical landscapes, and encourage visitors to interpret objects and spaces from individual perspectives.

Translating physical properties into the digital realm is a form of preserving cultural memory. However, this process often involves trade-offs that typically offer superficial approximations of real-world objects: despite advancements in gesture control, mixed reality, and haptics, digital representations, the loss of tangible qualities such as texture, weight, and materiality can challenge the authenticity of the translation. Our research works within these constraints, acknowledging that the true value of hybridised objects extends beyond their physical attributes into the abstract realm of cultural significance.

But such an act does not merely strip away the physicality of objects; it also reinterprets them, placing them within a new context where their value and meaning can be (re)assessed. Digitisation itself often reflects contemporary priorities around accessibility, preservation, and the global sharing of knowledge, but little has been done to examine how such a transformation might highlight (or obscure) the broader cultural values attached to artefacts and how this might be achieved in a responsible and ethical way. The research presented starts to unpack this issue, using corpus linguistics as a starting point to examine the multiplicity of meanings and understandings that relate to, and influence our understanding around artefacts and the embodied, physical world. This is a particularly important issue when we think about bringing the physical and digital together.

This includes accounting for issues around maintaining synchronicity between the physical and digital, and the need to retain meaning, authenticity, faith and cultural provenance. It also means considering aura, which can be defined as 'the feeling of respect, awe, and reverence that surrounds pieces of art as long as they are original pieces […] This concept has been received widely to describe a feeling of presence of art, persons, or objects' (Radermacher, 2019, p. 171-2 emphasis in original; see also Benjamin, 2019). If reproductions do not share the same aura as an original item, this has consequences for digital hybridity.

As Pietroni (2019) argues, '"communicating" the virtual is not a technological issue, but an epistemological question: technology cannot be fully effective without integrating the addressees' points of view and needs' (p. 12). This paper explores the concept of hybridisation, the blending of digital technologies and physical objects (cf. 'digital twins', virtual representations of a physical counterpart, which evolve over time to reflect changes in the physical system; see Wagg et al., 2024). As artefacts with representations across religious, secular, linguistic domains, the notion of relics provides a compelling case study for understanding how meaning and memory can be preserved, even as objects are translated into new forms. As the creation of relics is a cross-cultural and transhistorical human universal, they offer a superlative example with which to explore issues around hybridisation.

### *1.2. Religious relics and reliquaries*

Religious relics are objects of historical significance usually in the form of bodily remains or personal items belonging to saints or venerated individuals. Bodily remains can be whole bodies, or fragments of bone, skulls, flesh, blood, teeth, hair or fingernails, whilst belongings (or 'contact relics') can be clothing, shoes, cups, and weapons (Walsham, 2010). More rarely, relics may be perishable items. In Maoist China, mangos were assigned relic status (Walsham, 2010). While relics are not objects of worship per se, they serve as 'conduits of divine power' providing a link between earth and heaven and breaking down the barriers between the living and the dead (Razzall, 2020, p. 599).[1] Relics are significant in Christian, Islamic, and Buddhist faiths, believed to deliver miracles and spiritual comfort to believers. In Christianity, the veneration of saintly relics dates back to around 200 AD and became central to the faith by the early medieval period where they were used in public rituals, ceremonies, and individual acts of devotion, with people making pilgrimages to visit major relics of saints or martyrs.

Religious relics are usually housed within reliquaries—containers serving both a theological function by marking their contents as sacred, and a practical function by preserving objects from external elements (Razzall, 2020; Chaganti, 2008). By the mid-fifteenth century, reliquaries became conflated with the relics themselves (Montgomery, 1997), and in cases where a relic was lost, stolen, or destroyed, the reliquary often became the 'surrogate foci of devotion and reverence' (Razzall, 2020, p. 604) reflecting the fluidity of the boundaries between container and contained (Chaganti, 2008; Razzall, 2020). Like relics, reliquaries can take many forms. In early examples, saints' bodies were housed whole in casket-shaped reliquaries, but by the late Middle Ages, they took other shapes, whether they housed body parts or contact relics (Bynum & Gerson, 1997, p. 4). 'Speaking reliquaries' (so-called because their shapes 'speak' or express their contents) became particularly popular. These were shaped like body parts—arms, feet, heads—often to emulate their contents and convey the idea of fragmentation (Hahn, 1997; Bynum & Gerson, 1997). However, many speaking reliquaries did not contain the body parts they represented, reflecting a 'slippage between the contained and the container' (Hahn, 1997). One decorative form of reliquary is the 'ostensory' or 'monstrance', which displays a relic behind a crystal window; often featuring a sunburst design adorned with angels, they are used in The Eucharist or Holy Communion service. Reliquaries can also be diptychs or triptychs (two- or three-panel displays used as portable altars) or even simple or elaborately decorated boxes.

Reliquaries sometimes housed multiple relics within separate compartments (Bynum, 2017). There can also be multiple levels of containment, as with St Thérèse of Lisieux,[2] whose remains are protected by three containers: an alabaster box inside a wooden casket adorned with marble and gold, all within a glass case (Razzall, 2020, p. 598). All three containers are considered reliquaries, further underlining the complex relationship between container and contained. A pilgrims' journey towards a relic, through travel and progressively concealed spaces, is said to enhance their ultimate engagement with it (Turner & Turner, 1978). However, the final encounter is with the reliquary, which ultimately keeps the relic itself inaccessible to touch, and often to view.

The diverse and changing ways in which people have engaged with relics offer valuable insights into political, social, artistic, and religious developments (Bynum and Gerson, 1997). Most Medieval relics and reliquaries were lost or destroyed during the Reformation in the 1500s when Protestant reformers viewed the veneration of relics as acts of idolatry and superstition. The Reformation redefined relics as 'symbolic mementos' rather than objects of veneration (Walsham, 2010, p. 22); however, the Counter-Reformation which followed sparked a resurgence of interest in relics as items of religious power (Razzall, 2020). Today, relics may not be popularly viewed as the divine links they once were, but they often take on new significance as museum artefacts and objects of historical interest. For instance, the British Museum's 2011 exhibition, *Treasures of Heaven: Saints, Relics and Devotion in Medieval Europe* (Bagnoli, Klein, Mann & Robinson, 2011), showcased reliquaries of various shapes and sizes, including pendants, bottles, caskets, altars, and shrines. Serving as tangible links between the past and present, encounters with religious objects in museum settings can evoke ‘contemplation and a sense of divine presence’, revealing that the ‘tourist gaze is not necessarily devoid of religious feeling’ (Walsham, 2010, p. 33). This notion challenges the often-assumed secularity of museum spaces and signifies the complex, multifaceted nature of engagement with religious relics.

[1] Eastern holy images (icons) increasingly play the same role (Bynum and Gerson, 1997).

[2] A Carmelite nun, who was canonised in 1925 after her death from tuberculosis at age 24.

### *1.3. Relics beyond the religious*

The term 'relic' is polysemous, having transcended its original religious meaning to include various secular meanings. In line with this, Walsham (2010) describes the concept as having a ‘slippery, elastic, and expansive nature’ with boundaries that vary spatially, culturally and historically (p.11). In contemporary contexts, relics include grapho-relics, such as the signatures of secular heroes including film stars and musicians (Walsham, 2010) and artistic renderings from such artists as Monica Bock, who creates body-part ‘relics’ (Bynum & Gerson, 1997) and Daniel Arsham, whose ‘Relics in the Landscape’ exhibition features pieces hidden in woodland for people to discover (Yorkshire Sculpture Park 2014). The practice of ‘relicing,’ or artificially aging guitars, has been carried out since 1983 as a mark of respect to revered musicians, representing a hybrid of secular and ‘devotional’ elements (albeit not formally religious).[3] Early relic guitar prototypes were displayed behind glass at the 1995 National Association of Music Merchants show in a manner reminiscent of religious relics and their reliquaries. While not considered conduits of divine power, secular artefacts share similarities with religious relics in terms of their evocative nature and the connections they foster. As Walsham (2010) notes, relics, like ‘vestiges, fossils, and leftovers’, serve as reminders and memorials, creating a sense of identity and belonging (p. 13). Just as religious relics are considered ‘conduits of cultural memory’, tokens and mementos ‘serve as reminders and memorials and create senses of belonging and identity’ leading to some societies using the words more or less synonymously (Walsham. 2010 p.13).

Other senses of the term ‘relic’ go beyond the material. Linguistic relics, such as archaic words and verb forms, offer a window into the history and evolution of languages. English examples including archaic pronouns like ‘thou,’ irregular verb forms like ‘drank,’ and pronoun case markings as in ‘he’/’him’ serve as remnants of past linguistic practices, encapsulating broader cultural shifts over time (Culpeper, 2015). For example, the pervasiveness of the Anglo-Norman dialect lasted for several centuries and deeply influenced Modern English: more than half of the vocabulary that we use in the present day consists of words originally borrowed from Anglo-Norman including *exercise*, *guard*, *adventure*, *saint*, *salmon*, *vinegar* and *strange* (De Wilde, Rothwell & Trotter, 2021). Even the word *relic*, which ultimately derives from Latin *reliquiae* (‘that which remains after some process of reduction or elimination, the remnants’), uses a spelling that originated from Anglo-Norman. Living languages can be considered relics or depositories of the languages from which they derive.

The concept of ‘relics’ is considered central to the creation of historical dictionaries like the Anglo-Norman Dictionary (AND) and lexicography in general. For ‘dead’ languages like Anglo-Norman, no access to spoken word is available so a multifaceted world must be brought into being through physical documents that have been preserved over centuries. These are also considered to be relics of that language: remnants or objects that are left behind. However, like pilgrims and secular relics, dictionary-makers rarely engage with the original manuscripts; these precious, highly valuable items are hidden away inside libraries or churches to be protected and preserved. Instead, representations including photographs or digital images, always one or more steps removed from of the original relic, are all that may be available. These facsimiles preserve content as closely as possible. An editor who has read, transcribed, and interpreted a historical text presents it in the typescript of a modern book, often alongside a critical apparatus, a glossary, and an introduction. Sometimes a modern translation is made, adding another step removed from the original relic, whilst adding more meaning and significance to the text for the modern reader, who can better understand it. Any extraneous context added to present the relic to an audience can elevate, transform, obscure or even misunderstand it.

[3] Famous relic’d guitars include SRV#1, Stevie Ray Vaughan’s fender Stratocaster, which has been replicated by professionals and hobbyists alike; and Keith Richards’ ‘Micawber’ fender telecaster, which was specifically set up to be played with drop tuned slide guitar.

This leads us to focus on to the following research questions, prompting a deeper examination of both historical and contemporary texts to capture evolving interpretations of relics, in order that digitally mediated relics capture and preserve key information in cultural heritage contexts. In line with studies exploring the linguistic shifts in representations of money (Baker, 2011), the poor in seventeenth century England (McEnery & Baker 2017) and prostitution over the same period (McEnery & Helen Baker, 2016), we utilise historical and diachronic corpus linguistics to consider the following research questions:

1. What can Early Modern English and contemporary texts tell us about the (changing) status of relics and the values attached to them?
2. What lessons can be learned to navigate an ethically driven shift into digitisation, considering hybridity in relation to notions of faith, trust, autonomy and intent?
3. Can humanities-based methods, such as corpus linguistics, inform the digitisation and subsequent interpretations of cultural heritage?

We view these questions as catalysts for developing wider transdisciplinary understanding of these issues.

## 2. *Approach*

Corpus linguistics is a qualitative and quantitative approach allowing for the study of language through large collections of texts (corpora), using computational tools to analyse patterns, structures, and usage in real-world contexts. The discourse contained within corpora offers insights into identity, including the ideology of its producer(s). In a mutually reinforcing way, discourse both shapes and is shaped by the world around us. Additionally, it interacts with past and future discourse, the medium through which it is conveyed, and its intended and potential purposes (Eisenhart & Johnstone, 2008). Corpus linguistics allows for the examination of word frequencies and collocational behaviour (i.e., the habitual co-occurrence of words), along with an organised way to examine the wider context in concordance lines, which show a word or phrase of interest with its surroundings. Collocates can reveal patterns of positive and negative meaning and evaluation attached to words in a given discourse. This phenomenon is called discourse prosody, where a word takes on positive or negative meaning from its collocates. Semantic classification of collocates is typically supported by examining concordances (see Baker 2023: 143-146).

We explored the contemporary and historical use of the word ‘relic’ in order to better understand the prominence and social value of relics and track the development of ideas and beliefs across two time periods. Our specific focus is on modifier collocates since they can shed light on the status of relics and how they have been perceived and evaluated in contemporary and Early Modern English texts. In present-day English, the orthographic form of ‘relic’ and its derivatives (e.g., reliquary/reliquaries) are standardised, making analysis of a modern corpus of English quite straightforward. Historically, there is significant variation in spelling. We generated a list of search terms for historical corpus analysis with spelling variants sourced from The Oxford English Dictionary and EtymOnline (see 3.1). Using Sketch Engine corpus linguistic software (Kilgarriff et al., 2014), we identified statistically salient modifiers for the word *relic(s)* and its variants (e.g., 'sacred relic') before examining concordance lines qualitatively. The corpora for this study—Early Historical Books collection (EHB) and enTenTen21—were made available via Sketch Engine (Table 1); the English Historical Books corpus (EHB) comprises three smaller corpora: Early English Books Online (EEBO), Eighteenth Century Collections Online (ECCO) and Evans.[4] The corpora represent two different linguistic registers: books and web-sourced texts. While this approach is somewhat atypical, these corpora each represent the most representative language data available for their respected time periods. Early Modern perspectives are accessible only through written texts and the English Historical Books collection is one of the most comprehensive sources available, texts from which serve as proxies for historical viewpoints. Language sourced from the web in enTenTen21 offers a direct window into present-day attitudes and informal engagement that would not be captured by modern books alone.

Table 1. Corpora used in this study

| **Corpus name** | **Text date(s)** | **Size (tokens: words, punctuation, digits, abbreviations)** | **Language source** | **Corpus creation** |
|---|---|---|---|---|
| enTenTen21 | 2021 | 61,585,997,113 | English language web | Jakubíček et al. (2013) |
| Early Historical Books collection (EHB) | 1473 – 1800 | 987,242,247 | EHB comprises EEBO and ECCO, and Evans-TCP corpora as below (sub-corpora show in italic) | Text Creation Partnership (Evans-TCP created by Readex) |
| *- Eighteenth Century Collections Online (ECCO)* | *18th century* | *82,127,359* | *2,387 published works in English* | *PhiloLogic, part of the Early Historical Books collection* |
| *- Early English Books Online (EEBO)* | *1473–1820* | *792,281,394* | *English books* | *University of Oxford Text Archive, part of the Early Historical Books collection* |
| *- Evans-TCP* | *1639–1800* | *112,833,494* | *Evans-TCP corpus, which contains 5007 books published in America between* | *Text Creation Partnership (Evans-TCP created by Readex)* |

## 3. Results

### 3.1. Historical representations

To explore historical representations of 'relic', a list of modifiers was sourced from the EHB corpus. As spelling variations are not normalised for searches in this corpus, a search for the OED-sourced historical spelling variations performed using Corpus Query Language (CQL), limited to the 17 variations present in EHB: *relek*, *relic*, *relick*, *relicque*, *relik*, *relike*, *reliquar*, *relique*, *relicke*, *rellick*, *rellicke*, *relliek*, *rellike*, *rellique*, *relyke*, *relyque*, *rillik*.[5] There are 12,023 instances (12.18 per million words) of 'relic' and its variations (henceforth 'relic') in the EHB corpus: 10,521 (13.28 hits per million words) in EEBO; 1036 (12.61 hits per million words) in ECCO; and 466 (4.13 hits per million words) in Evans. Modifier collocates of 'relic' including adjectives, adverbs, nouns and possessives functioning as modifiers were extracted from the resulting extracts using 'Word Sketch grammar' in CQL (e.g., [ws("relic-n", "modifiers of \"%w\"", ".*")]. This search returned 3314 lines (excluding exact duplicates, i.e., text present in the corpus twice or more). The top 100 modifiers (excluding false hits, see Appendix A) were examined qualitatively and manually assigned to categories to determine broader patterns of meaning; these frequently denote evaluation, relational status, items, religion, condition, age, and cultural references (Table 2). An examination of (expanded) concordance lines aided classification of ambiguous modifiers including ‘noble’ and ‘pale’, which both refer to human remains (corpses or bones), and ‘rent’, which means ripped or torn in this context. The linguistic context for typical exemplars from each category is provided in Table 3.

---

[4] An open-source version of the EEBO corpus is available from the Text Creation Partnership https://www.english-corpora.org/eebo/.

[5] 23 are variants were not present in EHB:
raylect.*, realycke.*, releck.*, releick.*, releke.*, releqe.*, releque.*, relikke.*, reliqe.*, reliquik.*, rellyck.*, rellycke.*, rellyk.*, rellyque.*, relyc.*, relyck.*, relyk.*, relykke.*, relyqe.*, relyquyt.*, rerikes.*, rilik.*, rillyk.*

Table 2 - Top 100 modifiers of 'relic' (and variants) in the EHB Relics Concordance Corpus

| **Evaluation** | **Relational status** | **Item** | **Religion** | **Condition** | **Age** | **Culture** |
|---|---|---|---|---|---|---|
| precious / precyous / pretious | other | saints | holy / holie /holye | rotten | old | romish |
| small | such / suche | images | sacred | scattered | ancient | Columcille's |
| true | many | martyrs | popish | musty | new | |
| false | certaine / certain | dead | venerable | rent | | |
| superstitious | very | noble | religious | imperfect | | |
| sad | few | cold | blessed | poor / poore | | |
| dear | only / onely | choice | divine | | | |
| great | same | acid | apostles | | | |
| miserable | least | outcast | worship | | | |
| famous | several | crosses | hallowed | | | |
| counterfeit | last | church | idolatrous | | | |
| mournful | principal | ashes | | | | |
| goodly | like | pale | | | | |
| dear | divers /diuers | | | | | |
| infinite | various | | | | | |
| inestimable | more | | | | | |
| curious | little | | | | | |
| good | sole | | | | | |
| notable | sundry | | | | | |
| special / speciall | | | | | | |
| valuable | | | | | | |
| glorious | | | | | | |
| paltry | | | | | | |
| trifling | | | | | | |
| invaluable | | | | | | |
| wonderous | | | | | | |
| strange | | | | | | |
| wretched | | | | | | |
| louely | | | | | | |
| rare | | | | | | |
| strong | | | | | | |
| malignant | | | | | | |
| lamentable | | | | | | |
| meer | | | | | | |
| choicest | | | | | | |

## Table 3 - Example extracts containing modifiers of 'relic' in the EHB relics concordance corpus

| Reference | Classification | Modifier collocate | Example extracts |
|---|---|---|---|
| H01 | | true | pope may erre in canonizing your saintes, so may he much more erre in determining such and such reliques, to be the bodies, bones, or ashes of such and such saintes; and consequently, so may all papistes adoring them commit idolatry, yea though it were granted that **true reliques** might be adored; because as S. Austen grauely saide, their reliques are adored on earth, whose soules are broyling in hell fire. |
| H02 | | true | parts of their bodies are multiplied excedingly, (belike they were buried in a fruitfull ſoyle) and yet euery Prieſt, that ſheweth theſe reliques, commend that they haue to be the **true relique**, & ſome miracle, or viſion, euery one of them alſo would commonly faine , to proue that which they ſaid of their relique to be true. |
| H03 | | true | First, let vs knowe, that their Reliques are nothing else but forged deuices of their owne, and no **true Reliques** of Saints; as by one instance may appeare. For, the parts and parcels of wood, kept in Europe, which they say are parts of the crosse whereon Christ died, are so many, that if they were all gathered together, they would load a ship |
| H04 | | true | And s was the Church mocked (say I) by those which had reliques in them: not onely because they had no certaine rule of discerning **true reliques** from false; but also, because they were by such meanes induced, to place a speciall holinesse in those places, and led-on to the invocation of Saints departed. |
| H05 | | false | Idolatry is committed in the Church of Rome, by the worſhipping (at leaſt) of **falſe Reliques**, whereof there is ſuch a ſwarme: for the greateſt part of their Reliques being counterfait, the greateſt part of the worſhip which is done vnto them, muſt needes be Idolatry . |
| H07 | | counterfeit | Neuertheleſſe, we proteſt againſt the ſlanders of our aduerſaries, that albeit wee abhorre all falſe and **counterfait Reliques**, and refuſe to worſhip with adoration thoſe that are true: yet for theſe laſt ſort, when they are certainely knowne vnto vs, wee giue vnto them a due honour and reuerence |
| H08 | | counterfeit | I desire to know when false and **counterfeit Reliques** do work miracles, what we are to think of the Testimony given by such miracles and of the nature of them? It is a pleasant thing to see the accounts given by these men of the same Reliques being in several places at once. |
| H09 | | precious | It tooke this name, because in Saint Andrewes Church in Mantua, are kept as a most **precious relique**, certain droppes of our Sauiours blood; (thou canst not O Reader but belieue it) with a piece of the spunge. |
| H10 | Evaluation | precious | represented by the bodies of our Lord Iesus Christ, and S. Stephen: they vsually wrapped vp the most **precious reliques** of their Church in silken clothes, which they called (Sāctuaires) of holy coverleds as ecclesiasticall historiās write the which they haue since distributed, & sent abroad according to the anciēt custom of the church, |
| H11 | | precious | the grand Signior doth yearly send a Carpet or rather tombecloth of greene Velvet to cover the said Sepulcher, the old being then taken away and accounted the fees and vailes of their Priests and cleargie men that attend thereon, who cut the same into severall small pieces , and sell it to the superstitious at extreame rates for **precious reliques**~~; the Tombe it selfe being seated in a Temple built in Mecha, of no great magnificence or beautie, save the cost daily bestowed thereupon in Lampes of silver and gold~~ |
| H12 | | sad | It is a fire of hell that has blindnesse for smoke, scandal for light, and infamy and shame for ashes; these are the **sad relicks** of such, that having long prostituted themselvs, save of the ruins of their honor, nothing but a sad repentance |
| H14 | | miserable | originating in a ſpirit of intolerance which then prevailed, eſpecially in the Epiſcopal hierarchy; a ſpirit, now abhorred, as much by the enlightened and pious members of that communion, as perhaps by any others whatever; and which is viewed as a **miſerable relick** of ancient popiſh ſuperſtition and bigotry. |
| H15 | | miserable | from the Mauseoli, the Sepulchres, Temples, Villages, and Palaces, wherewith twas once proudly adorned on both sides, now only **miserable reliques** of its former lustre lying dejectedly and dispersedly in the waters. |
| H17 | | strong | ALL muſt of neceſſity have Original ſin or ſome Reliques of the Old man in them , yea ſuch **ſtrong Reliques** as will impell them to ſome Actual Sin or other, or to ſome tranſgreſſion of ſome of Gods Commandments |
| H18 | | goodly | Moreouer, forasmuch as I know you to be most deuout seruants of my Lord Barō S. Anthony, I wil shew you of my especiall grace and fauor, a most holy & **goodly relike** which my selfe brought long since frō beyond the seas out of the holy land, being one of the angel Gabriels fethers which he left in the virgin Maries chamber when he saluted her in Nazareth. |
| H19 | | mournful | Thus we have examined the peaceable and gentle ſpirit of the Engliſh epiſcopal church, and ſtill find in her the **mournful relicks** of her ancient perſecuting ſpirit |
| H20 | | lamentable | they would fain take from us this little vapour of honour that we have left, and these poor Titles, the unhappy remainders, and **lamentable Reliques** of the Italian Reputation, is a thing so hard to be digested, that by every honorable Baron of Italy it ought to be revenged; not with verbal complaints as I do, but with the daggers point. |
| H21 | | principal | Without all doubt the Law of nature, and the light of reason, was the rule whereby they were guided for the most part in such matters: which the wisdome of God would never have left in them or us, as a **principall relike** of his decayed image in us, if he had not meant, that we should make use of it, for the direction of our lives and actions thereby. |
| H22 | Relational status | diverse | a certaine priest of Rotenborough neare vnto Nicre, called Pfaff Isele, hauing bene at Rome and created the popes Indulgentiar, returned from thence into his Countrey furnished with **diverse reliques**, amonge others one of the winges of Michael the Archangel |
| H23 | | only | There are yet many excellent fountaines adorned with marble, and many arches, pillers, towers, ports and Temples, but most **only reliques** of lamentable ruines and sad desolation. |
| H24 | | only | I am totally unqualified to judge of its value: it bears, indeed, an high one to me—not on account of its own richneſs, but becauſe it is the **only relic** I retain of my deceaſed Father. |
| H25 | | sundry | in making some poore Saint (who (God knowes) meant simply and thought no hurt) beleeue that when he was liuing he had halfe a dozen heads, two or three dozen of eares, as many hands, and as |
| | | | many armes and legs: which imposture was sufficiently discouered aboue fifteene yeares ago, in a booke containing the Inuentory of **sundry relikes** of diuers countries. |
| H26 | | same | Sillery cut off a lock of his white hairs, which he begged I would preſerve for his ſake, and La Source gave me the **ſame relic**. |
| H27 | Religion | holy | For seeing that S. Chrysostome exhorteth the people not only to visit the Martyrs by repayring to their tombes, but also to touch, yea, and with faith to imbrace their reliques (for so hath the to haue thereby some benediction, doth he not plainely teach therein that **holy reliques** are to be reuerently kept, visited, and worshipped? |
| H28 | | holy | A Committee-man is the Reliques of Regal Government, but, like **Holy Reliques**, he out-bulks the Substance whereof he is a Remnant. |
| H29 | | holy | after the ſaid Battel finiſhed, and Victory atchieved, there was erected, and ſet up by the ſaid Prior and Monks, a fair Croſs of wood, in the ſame place, where they ſtanding with the **holy Relique**, made their prayers, in token , and remembrance of the holy Relique of St. Cuthbert, which they carried to the Battel. |
| H30 | | holy | They have also Altars which they anoint and consecrate ; & **holy Reliques**, whereof many doubtlesse are supposititious and false; therefore no new Reliques are to be received without the Bishops approbation, nor to be honoured without the Popes authority. |
| H31 | | saints | Her picture he adores twice a day, & for two howres together will not looke off it; a gar er or a bracelet of hers is more precious then any **Saints relique** |
| H32 | | saints | Oaths among the Papiſts are taken by touching the **Saints Reliques** that ſo the obligation of the oath may be divided betwixt God and the Saints. |
| H33 | | blessed | And many miriades of **blessed reliques** more & more encreased, and neuer fette the Popes blessing from Rome, for their warrant, so blessedly they multiplied, but had this Popes decrie bene plainly ment, or truly kept: bothe old and new , and all your blessed reliques, had bene banished from all blessednesse and worship, long agoe: as nothing but lies and forgeries, inuented to enrich your selues, with the spoyle of the peoples Idolatrie. |
| H34 | | popish | The Manna which putrified, being kept one day against Gods commandement, endured many hundred yeares by his appointment: but **popish reliques** are not privileged from putrifying, therefore God hath not ordained them so to be kept. |
| H35 | | idolatrous | Papists, e Iewes & Popes infinite traditions , aswel as these popish deuises & Iewish ordinances aboue-said; aswel as these popish **idolatrous reliques** of Fonte, Bells, Organs, Musick, Surplices, Coapes, Vestimentes, Habites, Hoodes, Cappes, Tippetts, Tires &c. |
| H37 | | hallowed | a Clerke (for ſo they call him) being farre in loue with a maid, and by no meanes either of long prayſes or large promiſes, able to gaine like affection at her hands, when he ſaw his hopes fruſtrate, and that he was not like to haue his purpoſe of her, turned his loue into rage, and cut off the maides head, which being afterwards hung vpon an Ewe tree, common people counted it as an **hallowed relique** till it was rotten |
| H38 | | venerable | Here are likewiſe the greateſt variety of **venerable relicks**, ſuch as St. Peter's beard, the ear of St. Francis, the milk of the virgin, with a thouſand fooleries beſides, all of which are in ſome ſenſe deified. |
| H39 | Item | dead | Pompey's faction, after they ſaw ſo many noble Senators (worthie in their iudgement, to haue beene honoured like gods, after death) deprived of all funerall rites and exequies; whileſt the **dead reliques** of meere carcaſſes, whileſt they lived, of paraſiticall mecanicks, or devoted inſtruments of tyrannicall luſt, were graced with Princely Monuments. |
| H40 | | noble | The **noble reliques** are pulled forth of their vnnoble ſepulchre. O thrice and four times, O bleſſed Apoſtles, what thankes ſhal we render vnto you, O who wil graunt vnto me to be rolled about the bodie of Paule, to be faſtened to his ſepulchre, to ſee the duſt of that bodie, fulfilling the thinges, which as yet were wanting in Chriſt, bearing the ſignes of his woundes. |
| H41 | | cold | When my aching heart ſhall have long ceaſed to beat, perhaps ſome kindred ſpirit may drop one tributary tear on my **cold relics**. |
| H42 | Culture | columcille's | Diarmaid, abbot of Aoi, went into Scotland, with **Columcille's reliques**. Diarmaid came into Ireland, with **Columcille's reliques**. |
| H43 | | romish | My good fathers & deare brethren, who ar first callid to the battel, to striue for gods glory and the edification of his people, against the **Romish reliques** and rages of Antichrist, I doubt not but that you wyl coragiouslye and constantlye in Christ, rape at these rages of Gods enemyes |
| H44 | | romish | Their Pomps, Rites, Laws and Traditions, are Antichristian, Carnal, Beggarly, Popish Fooleries; **Romish Reliques** , and Rags of Antichrist, Dregs and Remnants of Transform'd Popery. |
| H46 | Age | old | your praying to Saints and for the dead is in the purgatory already, your souls in the death; so your soules are lean, who feeds your fleshly minds with similitudes, images, Crosses, Crucifixes, **old Reliques**, pieces of cloath , or old bones, which you call reliques of Saints, so are full of dead mens bones |
| H47 | | new | Finally, that the Devill making vse of the superstition of these men, reuealed euery day **new reliques** , when he saw that the Bishops of the Romish church, were so earnestly giuen to enquire foorth the inventers and promoters of those reliques |
| H48 | | new | And heerein I commend their wiſedome, in choyſe of their reliques very much. Firſt, in that they tooke freſh greene **new reliques**, that were not antiquated, and out of date. For reliques (for oft wee ſee) worke like an Apothecaries potion, or new Ale: they haue beſt ſtrength, and verd at the firſt; and therefore Campians girdle, now like old Rubarb, begins to allay. |
| H49 | Condition | poor | What didst thou think to gain such a great Conquest over us, by thy beggarly Scraps and **poor Reliques**, that thou hast gather'd and learn'd under the Ministry of Presbyterians and Independants, when many of them (thy Tutors) have been confounded and overthrown, and their Mouths stopt by the Power of Truth in us, when they have gone about to overthrow it, and villifie us, by their perverse Gainsayings? |
| H50 | | poor | THE **poor reliques** of the caſtle are ſeated cloſe to the river; and are inſulated by a vaſt foſs cut through a deep bed of ſoft red ſtone |
| H51 | | imperfect | It hath been too great a fault in all ages, to wrap up their drugs in gold, and to vent false wares under glorious titles, imposing an the world, and on famous Authors many broken and imperfect Reliques. |
| H52 | | rent | And Ninivie, whose ruines are ruind: Seven ported Thebes, rich in silks and Furre, And Carthage, Africks glory, now declind: Nay, save of three, some monuments are showne, The other two, their seats, are hardly knowne. So Antioch, whence sprung the Christian name, And Sions Dame, Judeas |
| | | | sacred citie: Yea, Alexandria, famous in her fame, With Babylon, the remaindure of pittie: Though not like Jericho, a lumpe of stones, They're but **rent relicks**, of their former ones. |
| H53 | | rotten | Golgotha the place of skuls ſeemed to be deſigned on purpoſe, to upbraid and diſcourage our Redeemer; ſo many skuls and **rotten reliques** of humane frailty, as there were in that place, ſo many Trophies and Monuments of triumph did Death produce before the eyes of Chriſt |
| H54 | | rotten | the idolatrous inuocation and worshipping of Saints, the more then heathenish adoration of images & **rotten reliques** |

Modifiers of relic denoting **evaluation** include *true* (32 instances), *false* (30 instances), *counterfeit* (11 instances), *precious* (144 instances), *sad* (22 instances), *miserable* (17 instances), *idolatrous* (11 instances, see also Religion), *strong* (4 instances), *mournful* (7 instances), and *lamentable* (4 instances). Evaluative positioning is typically critical. Writers suggest that the canonisation of saints is prone to error and can lead to the idolatrous veneration of relics relating to unworthy individuals (H01) or express concerns over the Church having no way to distinguish between true and false relics (H04). 'False' and counterfeit relics are of significant interest (e.g., H05) and the credibility of miracles claimed to have occurred in connection with counterfeit relics is questioned, along with the dubious testimonies supporting them (H08). Some mock the veneration of relics by highlighting that many supposed pieces of the same body part relic exist in different places (H02) or that the excessive number of supposed fragments of the cross on which Christ died defies logic (H03). The writer of extract H09 doubts the validity of the relics held at Saint Andrew's Church in Mantua (drops of Christ's blood and a piece of the crucifixion sponge), using sarcasm—'thou cans't not O reader but believe it'—to challenge the credulity needed to accept them as genuine. Many extracts are concerned with commercialism. For example, H11 discusses the annual replacement of velvet cloth surrounding the sepulchre of a significant religious figure in Mecca. For this writer, the practice of selling pieces of the old cloth as relics to 'superstitious' people raises concerns about the exploitation of faith. A non-critical view is represented in a minority of texts. Some describe the significance of relics as connections to holy figures and events (e.g., H18) or affirm that the veneration of relics, when practiced correctly, involves honour and reverence, rather than adoration (H07). Extract H10 recounts how relics associated with Jesus Christ and Saint Stephen were wrapped in fine silken cloths (called *Sanctuaires* or holy coverings) and then distributed to different churches.

The term 'relic' is also used metaphorically. For example, relics are the remnants of a life lived in moral degradation: regret and the ruins of former honour (H12); the remnants of sinful nature inherent in humans ('strong relics'), which cause them to commit sinful transgressions (H17); and the remnants of intolerance associated with Catholic 'superstition and bigotry' (H14). Similarly, 'mournful relics' (H19) relates to historical 'persecuting spirit' of the English Episcopal church and the influence this continued to have on its identity (despite later being considered peaceful), and 'lamentable relics' describes a diminished national reputation (H20). References to non-religious material things are also present in these extracts. For example, the term 'miserable relics' communicates feelings of loss and nostalgia for significant buildings and structures that have fallen into disrepair (H15).

Modifiers denoting **relations** of quantity or comparison including *principal* (10 instances), *diverse* (14 instances), *only* (17 instances), *sundry* (4 instances), *same* (12 instances), contain further criticism. For example, H25 highlights the miraculous or absurd attributes given to some saints posthumously, while H22 describes a 'certain' priest's return from Rome with diverse relics including the wing of Archangel Michael. The theme of moral and physical decay is also present here in relation to the decline of divine image (reason and morality) in humanity (H21) and structures that have been reduced to 'lamentable ruines' (H23). Finally, relic status is attributed to mementos including objects belonging to the deceased loved ones (H24) and locks of hair exchanged as tokens (H26), reflecting the blurred boundary between relics and tokens.

Modifiers denoting **religion** include *holy* (395 instances), *saints* (43 instances), *blessed* (8 instances), *popish* (29 instances), *idolatrous* (8 instances), *hallowed* (7 instances), and *venerable* (17 instances). Relics are again heavily criticised here, particularly in terms of authenticity. For example, relics such as St Peter's beard, the ear of St Francis and the milk of the virgin Mary are described as 'fooleries' (H38) and the sacredness of deteriorated relics is questioned by analogy with manna (food provided by god to the Israelites), which was both preserved and decayed at god's will (H34). Many consider relics to be fraudulent and exploitative. Through repeated use of 'blessed' ('so blessedly they multiplied'), the writer of H33 cynically implies that relics are not truly divine but a product of human invention and greed. Practices surrounding relics are highly criticised. For example, invoking saints' relics in oaths by 'Papists' (a pejorative term for Roman Catholics) is said to complicate the obligation they have to god (H32); and 'Popish' relics including fonts, music, instruments and clothing are said to detract from Catholic and Jewish faiths (H55). Relatively few religious references describe relics in a positive light. Extract H27 discusses St. Chrysostom's teachings in support of venerating relics, whilst H30 emphasises caution in accepting new relics to ensure that only authentic relics are honoured. The intertwining of faith and warfare features in a description of a wooden cross erected at a battle site to commemorate victory and honour the holy relic of St. Cuthbert, faith in which is believed to have played a role in the victory (H29). Secular objects tied to romantic or tragic love are often elevated to the status of holy relics. For example, H31 describes an individual's devotion to a woman, describing her belongings as more precious than any Saint's relic, whilst H37 relays the story of a clergyman who killed a maid after his unrequited love turns to rage. People treated her decapitated head as a 'hallowed relic', suggesting that they attributed religious or cultural significance to it (albeit only until it decayed). Finally, metaphorical use is present in these extracts, as illustrated by H28, which likens the role of the 'committee-man' to holy relics, implying that the role is a remnant of a bygone regal governance system, a part of something that once held substantial power.

Modifiers denoting **items**, *dead* (12 instances), *noble* (10 instances) and *cold* (7 instances), all relate to bodies. Worthiness is a strong pattern of representation here. For example, extract H39 uses the term 'dead reliques', to imply that the physical remains—'mere carcases [meere carcaſses]'—of individuals aligned with an opposing Roman faction were given undue honour, whilst extract H40 describes the remains ('noble relics') of apostles being taken from a tomb (sepulchre) that was not considered worthy of their significance. The writer of extract H41 uses the term 'cold relics' to describe their own body after death, expressing the desire for remembrance by a close person.

Modifiers denoting **cultural references** include *Columcille* (4 instances) and *romish* (26 instances). For example, extract H42 describes the journeys of an abbot as he travels to Scotland and Ireland carrying the relics of Colmcille (Columba), a canonised Irish abbot. The act of travelling between places carrying relics strengthened ties between regions through shared reverence. The writer of H43 rallies fellow clergy and believers to engage in a spiritual battle against the idolatrous practices of venerating of relics by Catholic 'enemies of god'. This extract (H43) and H44 use the modifier 'Romish' in the same was as 'popish' above to undermine Catholic practices, including describing relics as 'fooleries', 'dregs', and 'rags of the antichrist' (H44).

Modifiers denoting **age**—*old* (22 instances) and *new* (12 instances)—provide mixed perspectives. Relics are dismissed as pieces of cloth and old bones (H46) and the continual introduction of new relics is considered exploitation by the devil to deceive superstitious people (H47). Conversely, the writer of H48 praises the discernment in choosing relics, emphasising the value of newer relics over older, less effective ones; here, relics are likened to fresh potions and ale to illustrate that their powers are strongest when new.

Lastly, modifiers denoting **condition** including *poor* (21 instances), *imperfect* (6 instances), *rent* (5 instances), and *rotten* (18 instances) echo many of the same themes above. The practices of invoking saints, adoring images, and venerating relics are characterised as idolatrous and more extreme than pagan practices (H54); and the writer of H49 confronts those who attempt to assert superiority using 'poor relics' sourced from discredited Presbyterian and Independent 'tutors'. 'Poor relics' and 'rent relics' describe the dilapidated remains of a castle (H50), and to reflect on the decline of once-great cities like Nineveh, Thebes, Carthage, Antioch, and Alexandria (H52). Extract H51 discusses the deception of presenting inferior goods ('drugs') as valuable through attractive packaging and grand titles. It is not clear whether 'relics' refers to religious items here—though Reformist criticism about packaging could certainly apply to reliquaries—or whether it refers more broadly to items that are falsely attributed great value or significance. Finally, H53 submits that the presence of skulls and decaying remains ('rotten relics') at Golgotha (Calvary, the site of Jesus' crucifixion) was deliberately designed to mock and discourage Christ during his crucifixion.

### *3.2. Contemporary representations*

Moving on to exploring contemporary representations of 'relic', we examined the top 100 modifiers of relic extracted from enTenTen21. These modifiers were identified using Sketch Engine's 'Word Sketch' tool, which reveals collocations and word combinations, with results separated by grammatical relationship to the search term. The modifiers, broadly classified by meaning, frequently denote religion, age, condition, relational status, value, and evaluation of the character of relics (Table 4). Ambiguous terms for which wider context was needed to establish meaning include: 'barbarous', which relates to a gold standard; 'sacred', which relates both to religion and the video game Halo; 'forerunner', which also relates to Halo; 'empress' and 'desolate [bones]' from works of fiction, 'ayleid' from the video game Elder scrolls; and 'lacon' from the media franchise Transformers. The linguistic context for selected exemplars from each category is provided in Table 5.

Table 4 - Top 100 modifiers of relic in the enTenTen21 corpus

| **Evaluation** | **Age** | **Item** | **Religion** | **Culture / Media** | **Relational Status** | **War** | **Condition** | **Classification** | **Politics** |
|---|---|---|---|---|---|---|---|---|---|
| barbarous | ancient | tooth | holy | cultural | forgotten | cold | dusty | first-class | Crow |
| priceless | prehistoric | archaeological | sacred | forerunner | stolen | WWI | rusty | second-class | Jacobite |
| precious | Jacobite | artifact | demigod | sacred | immovable | war | well-preserved | third-class | |
| quaint | historical | relic | incorrupt | desolate | lost | battlefield | fading | | |
| barbaric | archaic | architectural | buddha | Ayleid | long-lost | wartime | | | |
| obsolete | historic | dimensional | Buddhist | lacon | | | | | |
| outdated | Roman | statue | god | museum | | | | | |
| venerable | colonial | bone | saintly | empress | | | | | |
| prized | feudal | ruin | saint | | | | | | |
| antiquated | age | shipwreck | sacred | | | | | | |
| curious | Egyptian | sword | passion | | | | | | |
| anachronistic | Mayan | unholy | holy | | | | | | |
| useless | antiquarian | miracle-working | devotional | | | | | | |
| cursed | dynasty | miracle-working | religious | | | | | | |
| valuable | medieval | | | | | | | | |
| wonderworking | era | | | | | | | | |
| cherished | bygone | | | | | | | | |
| mystical | | | | | | | | | |
| mysterious | | | | | | | | | |
| outmoded | | | | | | | | | |
| irrelevant | | | | | | | | | |
| magical | | | | | | | | | |
| fascinating | | | | | | | | | |
| irreplaceable | | | | | | | | | |
| antique | | | | | | | | | |
| unholy | | | | | | | | | |
| authentic | | | | | | | | | |
| interesting | | | | | | | | | |
| miracle-working | | | | | | | | | |
| embarrassing | | | | | | | | | |

## Table 5 – Example extracts containing modifiers of ‘relic’ in enTenTen21

| Reference | Classification | Modifier collocate | Example extracts |
|---|---|---|---|
| C01 | Items | tooth | they were my daughter's **teeth, relics** from my days as the Tooth Fairy. |
| C02 | | | It is reputed to hold a holy **tooth relic** given to Anawratha by the King of Sri Lanka. |
| C03 | | | explore its old colonial Garrison cemetery and see the famous **tooth relic** from Buddha at the Temple of the Tooth. |
| C04 | | bone | Famen Temple is also renowned for the finger **bone relic** of the Sakyamuni Buddha preserved in the underground palace. |
| C05 | | | It was for the first time that DNA testing was done on a **bone relic** in India at CCMB, Hyderabad. |
| C06 | | architectural | The word ruin, as most **architectural relics** are called, has its origin in a Latin word ruina from ruere, which is 'to fall’. |
| C07 | | | This palace is the largest and best kept **architectural relic** of that period, with countless mural paintings, carving and other precious art crafts. |
| C08 | Religion | holy | *Joseph brought the **holy relic** to Britain where it was eventually concealed in a mysterious castle surrounded by a blighted landscape.* |
| C09 | | | *four priests carried the **holy** relics of Saint RAPHAEL in a procession around the monastery church* |
| C10 | | devotional | *these relics are a piece of scalp hair or a piece of wood, which are allowed by eBay policy. They are sacred and **devotional** relics of the Church. The item will be shipped from Vatican City by DHL(delivery worldwide in 2-3 days)* |
| C11 | | saintly | *They wanted to divide Charbel's body and bones among them, as **saintly** relics, which they intended to wear on their person and place in their homes* |
| C12 | | | *Leominster's early history as an important religious centre and repository of **saintly** relics is explored* |
| C13 | Evaluation | useless | *I bought a Synology NAS and now am looking at Dropbox as a **useless** relic.* |
| C14 | | | *Tonsils were routinely cut out of children's throats under the assumption they were **useless** relics of our evolutionary past* |
| C15 | | fascinating | *visitors can explore **fascinating** relics of the lost Gaelic kingdom of Dál Riata* |
| C16 | | | *Its great castle is a fascinating **relic** of the mysterious Templars, full of secret passages, symbolic paintings, and compelling art.* |
| C17 | | cherished | *Today that crucifix is kept as a **cherished** relic that continues to influence modern day commanders of the 1-69th.* |
| C18 | | | *This Victorian market is a **cherished** relic of times gone by.* |
| C19 | culture | cultural | *Many priceless artworks and **cultural relics** were taken out before being burned.* |
| C20 | | | *These have included 19th century masterpiece paintings from France, ancient **cultural relics** from China, and even a set of dinosaur fossils.* |
| C21 | | | *Hubei Provincial Museum in Wuhan, home to some of China's oldest **cultural relics*** |
| C22 | | museum | *Today the abacus and the gold scale are **museum relics**, reminders of the traditional employment of whatever was best and most useful at any period of time.* |
| C23 | | | *Some may remember the plastic slip-covers that protected the furniture that sat like a **museum relic** to be viewed but not touched.* |
| C24 | | | *Most similar engines that survived to become **museum relics**, he said, were refitted again and again over decades, and represent hybrids with modernized parts.* |
| C25 | Age | Jacobite | *One of a collection of **Jacobite relics** amassed by Sir John Hynde Cotton and his descendants - oval miniature in enamel of George II* |
| C26 | | | *This work was published in "Hoggs **Jacobite Relics**.”* |
| C27 | | historical | *One of the important **historical relics** of the village is the Monastery of Mar Odisho* |
| C28 | | | *However, viewing statues as **historical relics** and primary sources in their own right leads one to approach the topic with gravity* |
| C29 | | | *Anarchism, which was considered a quaint **historical relic** at best, rose from the grave to become the West's most vibrant political movement.* |
| C30 | | roman | *Many Roman coins and other **Roman relics** have been found in Alderney.* |
| C31 | | | *Interesting fossils and **Roman relics** have been found in and near the cave.* |
| C32 | Status | stolen | *her path has brought her here to Arkham, following a shipment of **stolen relics*** |
| C33 | | | *A piece of art he purchased in 1985 was actually a **stolen relic** from a church in Cyprus.* |
| C34 | | lost | *worked tirelessly for over 20 years, seeking out **lost** ancient **relics** of the natural world, and rescuing at least 270 varieties from obscurity.* |
| C35 | | | *Players will seek out **lost relics**, outwit terrific beasts, and venture into the entrancing Otherworld of the Fae.* |
| C36 | | | ***LOST RELICS*** *OF THE KNIGHTS TEMPLAR had its U.S. premiere on Wednesday, November 11, 2020 at 10:08pm ET/PT on Discovery Channel.* |
| C37 | War | battlefield | *The store is best known for its Civil War memorabilia, which can range from* ***battlefield relics*** *and artifacts, to books and documents.* |
| C38 | | | *collectors are finding* ***battlefield relics*** *on a daily basis, actually in the ground or in Depot Ventes and Brocantes.* |
| C39 | | cold | *Are U.S. nuclear weapons merely* ***Cold*** *War* ***relics*** *that belong in "the dustbin of history" along with communism and the Soviet Union* |
| C40 | | | *many wonder why Bond is still a thing, given that he's both a* ***Cold*** *War* ***relic*** *and an outdated symbol of masculinity.* |
| C41 | | WWII | *There are fading photographs and numerous* ***WWII relics****, including massive guns, torpedoes and a corroding Zero fighter. The centre is open daily and charges K5 for entry.* |
| C42 | | | *He couldn't believe that the person on the phone had a friend in France who found his dog tags while searching for* ***WWII relics*** *with a metal detector in Normandy.* |
| C43 | Politics | crow | *Obama's phrase, "****Jim Crow relic****," has been adopted by activists and elected Democrats all the way up to President Biden.* |
| C44 | | | *The Senate filibuster is a* ***Jim Crow relic****, historically used to protect the South's dependence on slave labor and later to defend segregation and block civil* |
| C45 | | | *'Right to work is a* ***Jim Crow relic*** *that was specifically designed to keep White and Black workers apart, playing on our worst fears* |
| C46 | Condition | well-preserved | *This* ***well-preserved relic*** *from the times of Ancient Rome allows you to see the original bathhouse from thousands of years ago.* |
| C47 | | | *Sotheby's presents a rare example of a remarkably* ***well-preserved relic*** *from over 3,000 years ago.* |
| C48 | | rusty | *Peterson wanted to show me the last remnants of the last active whaling station in North America – a* ***rusty relic*** *of the area's past.* |
| C49 | | | *Our blogs are filled with beautiful landscape photographs, nature and even the odd* ***rusty relic****, so there is something here for everyone* |
| C50 | | dusty | *The Ultima is a modern car, not some* ***dusty*** *'90s* ***relic*** |
| C51 | | | *To our great-grandchildren, the horror of the Holocaust may become a* ***dusty relic*** *of antique memory, much as the Spanish Inquisition is to us.* |
| C52 | | fading | *the Gran Hotel París, a* ***fading relic*** *of yesterday's glories that beams out the word "París" across the city skyline after dark.* |
| C53 | | | *Their species of Southern Democrat will be, by then, a* ***fading relic*** *of a strange, distant, and inexplicable past.* |
| C54 | | | *Some regard national sovereignty as a* ***fading relic*** *of a pre-globalized world.* |
| C55 | Classification | first-class | *There are three classes of relics: A* ***first-class relic*** *is a piece of the body of the saint.* |
| C56 | | | *the National Shrine of Our Lady of the Miraculous Medal with an heirloom,* ***first-class relic*** *of St. Catherine Labouré.* |
| C57 | | | *The shrine is a place of pilgrimage housing two* ***first-class relics*** *of St. John Paul II.* |
| C58 | | second-class | *A* ***second-class relic*** *is an item that was owned or used by the saint.* |
| C59 | | | ***Second-class relics*** *are something that a saint personally owned, such as a shirt or book (or fragments of those items).* |
| C60 | | | *The legitimate acquisition of a* ***second-class relic*** *of Blessed Frassati (like the one pictured here) is possible but the process is lengthy* |
| C61 | | third-class | *A* ***third-class relic*** *is an object that has touched the tomb or first-class relic of the saint.* |
| C62 | | | *touched the reliquary with rosaries, holy cards or other religious items, which turned them into* ***third-class relics*** |
| C63 | | | ***Third-class relics*** *are objects which have been touched to a* ***second-class relic****.* |

Modifiers relating to **religion** in enTenTen21 include *holy* (5680 instances); *devotional* (83 instances, almost all of which are from eBay listings); and *saintly* (126 instances). These religious relics are something to share stories or accounts of (C08, C09, C11), or something sacred to visit (extract C12), possess or own (C10, C11). Perhaps surprisingly, modifiers of 'relic' denoting **items** are not typically religious, though they do include teeth ('tooth', 564 instances); bones (293 instances); architecture ('architectural', 368 instances). In this context, sacred teeth and bone relics are something to touch, see, test, or be near (C02, C03, C04, C05) but teeth also appear in non-secular context as sentimental objects saved from a childhood (C01). Architectural relics might either be ruins (C06) or well-preserved buildings constructed in a discontinued style (C07). Modifiers denoting the **classification** of relics represent a significant part of the way they are considered in modern texts (179 instances of *first-class*; 64 instances of *second-class*; and 53 instances of *third-class*). Although this kind of classification is being phased out and has not been used by the Vatican since 2017, it remained part of UK English web discourse in 2021 (e.g., C55, C56). Interestingly, the new classification descriptors, ‘significant’ and ‘non-significant’, do not appear in the top 100 modifiers examined (for details see The Vatican, 2017).

Modifiers signifying the writer’s **evaluation** of the relic being discussed include *useless* (165 instances); *fascinating* (273 instances); and *cherished* (96 instances). Evaluative positioning may be negative in descriptions of obsolete things such as technology that is falling out of use in place of a superior alternative (C13) or body parts (e.g., tonsils) assumed to have no necessary function (C14). Finally, evaluative modifiers may also promote items to tourists (C15, C16); they may describe a prized item which holds considerable influence (C17), or an historical place that maintains significant cultural value (C18). Modifiers denoting **culture**, including *cultural* (6842 instances) and *museum* (189 instances), appear in relation to something of high cultural value that has been destroyed (C19), and very often appear in reference to Chinese items that have either been removed from their origin (C20) or to items that are an attraction for visitors to a place of interest (C21). The use of the modifier ‘museum’ is largely metaphorical and not used for items that are housed in a museum; like ‘useless relic’, a ‘museum relic’ is an (assumed to be) obsolete item.

Modifiers of ‘relic’ frequently denote **age**, which may reference a specific time period—for instance *Jacobite* (122 instances) and *Roman* (165 instances)—or more may be general—*historical* (3252 instances). Relics defined by their age are part of collections (C25), significant buildings (C27) or primary sources for study (C28), or items of historically significant interest that have been found (C30, C31). Other contexts relate to a revived political movement (C29) or songs about a political movement (the Jacobite risings) published in a volume (C26). Just one other modifier, ‘crow’, related to **politics** (70 instances). Like ‘Cold War relics’, ‘Jim Crow’ relics (C43, C44, C45) are historical political ideas, activities or conducts that continue to be used even though they are considered outdated.

Modifiers denoting **war** include *battlefield* (81 instances); *cold* [war] (470 instances); and *WWII* (127 instances). Battlefield relics (C37, C38) and WWII relics (C41, C42) are usually framed as items of historical interest, whereas Cold War relics (C39, C40) are outmoded things that (the writer believes) should be forgotten.

Modifiers denoting the **relational status** of relics describe their position in relation to people. For instance, *stolen* (213 instances) and *lost* (210 instances). These are usually items to seek (C32, C34, C35, C36) but more rarely they may refer to a rediscovered item (C33). If the modifiers denoting the **condition** of relics are positive (e.g., *well-preserved*; 73 instances), they refer to genuine historical items, which are interesting in function or appearance (e.g., C46, C47). If the modifiers denote poor condition, such as *rusty* (150 instances), *dusty* (317 instances), *fading* (49 instances), they are typically used metaphorically to describe unfavourable and outdated things such as a region’s outlawed past (C48); an undesirable vehicle (C50); an event that has (regrettably) diminished in cultural memory (C51); or a political concept from the past (C54), with the implication that it may not be appropriate for contemporary times. Less often, however, these modifiers refer to what the writer considers to be a positive change, such as the disappearance of people holding certain political beliefs (C53); a symbol of a better time (C52); or vehicles featured in a travel blog to attract readers (C49).

### *3.3. Summary of findings*

In the Early Modern period of English, relics serve as multifaceted symbols—tools of religious and political control, objects of moral and spiritual significance, and as commodities representing religious materialism. Much criticism highlights the complex relationship between belief, power and identity. In debates about morality, relics are representative of deeper struggles between tradition and reform and form the basis of discussions about authenticity, authority, and faith during a period of profound change. Believers attributed real power to relics leading to greater demand. Relics have been commoditised in all cultures (see Walsham, 2010) but the corpus analysis highlighted significant unease about commercial exploitation in religion and scepticism over relic authenticity. Criticism often comes from a Protestant perspective, which framed relics as instruments of Catholic control, deception, and exploitation. The outgrouping of those who venerated relics reflects deep social divisions. By positioning relic worshipers as followers of 'pseudo-Christianity', a division between genuine belief and superstition is created and used as a way of distinguishing between 'true' and 'false' Christians.

Tokens and mementos are described as relics in both corpora. These items are not always easily distinguishable from religious relics in that they each serve as memorials and are closely tied with belonging and identity, and these blurred boundaries make it difficult to disentangle meaning (Walsham, 2010). The framing of relics as items of religious significance and desirable objects to visit or value plays a lesser role in contemporary texts, though it does still appear. Relics are more typically framed as symbols of cultural heritage, representing past events, places, and traditions in contemporary language use. When a historical period is idealised or romanticised in contemporary texts, related artefacts—such as 'WWII relics' or 'battlefield relics'—become items to be sought out, admired and treasured. Conversely, when an historical period is viewed negatively, associated 'relics' are framed pejoratively, referring to archaic, obsolete ideas rather than physical objects. The word 'relic' can take on negative meaning (negative semantic prosody) from the modifiers it habitually co-occurs with (e.g., *useless*, *fading*, *dusty*, *rusty*) in the contexts of describing secular items such as cars or used in some other metaphorical sense. Finally, the contemporary texts revealed a shift from objects of physical significance to those of digital significance. Whilst descriptions of human-remains such as teeth and bones, typically refer to physical items, man-made items such as swords, were referenced more in digital contexts, particularly computer gaming. Though this analysis focused on modifier collocates, the examination of (expanded) concordances allowed for a broader examination of linguistic context. Modifier collocates are just one lens through which to examine this topic; other collocates may reveal additional themes. As the representation of Catholics have emerged from these concordance lines, an in-depth analysis of social actors surrounding relics may also prove fruitful. Modifiers sometimes encompassed multiple meanings but were semantically classified according to the most dominant usage in line with the wider context; that said, patterns appearing across the classifications were of primary interest here. We aimed to capture a broad understanding of major representations of relics as a catalyst for discussion on the kinds of issues that would need to be considered in digital hybrid design. A more fine-grained analysis could be done in future to better understand change across the early modern period of English as well as differences across discourse genres.

### ***4. Discussion***

The creation of digital relics could allow institutions to continue their religious, educational, and cultural missions even when physical access is restricted (see Curtis 2019; Simone et al., 2021). Having presented insights into the different meanings and perspectives surrounding the word 'relic' in historical and contemporary language use, we now explore what lessons can be learned to navigate an ethically driven shift into digitisation, considering hybridity in relation to notions of faith, trust, autonomy, and intent, framing our discussion from transformative (i.e., part to whole, specific to general); transdisciplinary (what the findings mean for hybridisation); and translational (products, practices and uses) perspectives (Cao, 2023). More than this the research presented, which is based around a single term in our case, theoretically and practically demonstrates the need to appreciate that using Humanities based approaches can offer technologists a different lens through which to start to think about design. This is particularly important when we think about hybrid technologies, which embody elements of the digital and physical. As we look towards developing and applying AI to such technologies we need to be aware of the ways that language influences our understanding, engagement, and provides a mechanism through which we can interpret and interact with intelligent artefacts. Of course, as our research has shown, this is in some ways problematic for designers and technologists as language and culture aren't static, and represents a multitude of times, places and reasoning about the world. In many respects the Humanities as research area is aware of this, but we see this piece as a translational work that enables researchers beyond the Humanities a route to start to think and theorise about such issues. In the next sections we discuss the transformative, transdisciplinary and translational nature of the work to show the theoretical underpinnings of the research.

### *4.1. Transformative*

The language we have examined is a representation, a way to discuss feelings, thoughts, and beliefs that is very much a 'reduction' of the society and physical world it communicates. Our lens into that world is inevitably shaped by the written texts that remain. Language surrounding religious and secular relics offer us a point of reflection into the lived experience of the actors involved. The historical texts explored here have been digitised and processed for corpus linguistic analysis as part of the EHB corpus. This act simultaneously increases access to the content to any interested researcher and increases our distance from the original, physical books. Any decisions made along the way (e.g., editorial decisions or choices about what metadata to capture) runs the risk of compromising the material substance of the original texts. Such considerations hold true for the hybrid transformation of relics and artefacts, which also represents a departure from the original object and must account for the various perspectives shaping understanding of these objects. While this endeavour necessarily involves a degree of interpretation, historical linguistic evidence can be contrasted with an understanding of contemporary perspectives to communicate meaning in a way that resonates with users of hybrid digital technologies.

The linguistic representations considered above offer a fragmentary perspective both because the texts are from snapshots in time and because the texts are part of, and shaped by, wider, interconnected discourses. Stable patterns in representation do appear across historical and contemporary corpora, however. For example, relics carry meaning beyond their physical form, functioning as both individual artefacts and integral parts of broader cultural and religious narratives. Relics are 'inherently paradoxical, being fragmentary and yet also complete' (Razzall, 2020, p. 604) and digitisation must recognise that they are not isolated artefacts, but part of a larger 'whole' including cultural traditions, belief systems, and historical narratives, which will influence how users perceive faith and trust in the digital realm.

Rendering the relic digital transforms the computer into a modern reliquary, removing the visual barriers arising from spatial distance and the concealment of sacred objects within reliquaries, whilst offering (perhaps the ultimate) protection against decay and damage. Hybrid relics can serve to increase access, but such an endeavour can only hold value if this is done in an ethically responsible way, which accounts for value gained and lost in the process, according to a range of perspectives. While the tangible significance of the original relic and its physical container are undoubtedly lost, this shift may hold less weight for modern audiences given that secularity is not a dominant part of the way relics are presented in contemporary texts.

### *4.2. Transdisciplinary*

The findings have significant implications for hybridisation in relation to shaping and reflecting diverse perspectives, which have differed both within and across time periods. Just as we have only been able to examine the perspectives of those with discursive 'power' (Fairclough, 2013) in the linguistic analysis, we must recognise that those with the power to enable hybridisation have responsibility to balance historical and modern understandings of relics. When a digital version of an artefact is created, metadata captures and preserves key information. Decision-making around whether certain historical or contemporary perspectives are foregrounded over others (e.g., how to balance religious tensions sensitively); and whether to foreground historical (where the aim is preservation of cultural heritage) or contemporary perspectives are key in this endeavour. Much will depend of course on the context of the digital representation and the primary intended audience.

Unlike digital twinning, which mirrors real-word counterparts in real time, hybridisation creates a new layer of representation—a 'future relic' capturing a representation from a moment of time. Hybridisation should focus on creating nuanced, immersive digital experiences that preserve the authenticity and agency that individuals would have with physical relics. Just as interactions with physical relics can feed into shared narratives around trust in authenticity, individual digital interactions have the capacity to influence collective trust. As such, audiences engaging with relics in digital spaces must retain agency over the process to mirror autonomy over engagement with faith practices. Done well, hybridisation will enable broader accessibility and (re)interpretation of relics' meanings but the endeavour requires careful attention to the immersive and sensory dimensions of engagement, which are essential to human interactions with cultural artefacts (Pietroni, 2019). To account for multiple instances of the same relic, Medieval churchmen theorised that they self-generated like the loaves and fishes that fed the five thousand (Walsham, 2010, p. 12). But as the Early Modern texts illustrate, much scepticism surrounded 'duplications' of the same relic. In the same way, reproductive technologies can 'diminish the aura' surrounding sacred objects (Walsham, 2010, p. 12). As such, issues surrounding the security of digital assets and protecting against falsification, modification, spread and fragmentation of data (Wang et al., 2021) are particularly important. Though these risks can compromise the educational and cultural mission of museums, solutions such as digital watermarking and blockchain data management may offer solutions (Trček, 2022; Wang et al., 2021).

### *4.3. Translational*

The 'tension between closeness and distance' characterising interactions between Medieval pilgrims and relics (Razzall, 2020, p. 602) is echoed in both Early Modern English and contemporary texts, though more so in relation to non-secular items in the modern corpus. We must consider how resolving this tension by removing material occlusion (i.e., making the relic fully visible) might affect such interactions. Whether approaches such as those used in the VRtefacts project—where 3D-printed 'relics' can be touched and VR versions viewed—enhance or diminish the sacred experience of relics in the present day is yet to be explored. Digital technologies have the potential to both reflect and augment (experiences of) objects carrying personal, cultural and religious significance. Technologies that may facilitate hybridisation of relics include photogrammetry, which enables the photorealistic capture of physical objects, rendering them as a 3D graphics which can be explored in a 360º viewer; and AI driven optical character recognition (OCR), which can learn, analyse, and transcribe physical inscriptions such as medieval texts into digital documents. For those working in museum experience design, the 'priority objectives' as outlined by Pietroni (2019, p. 2) include: 'improvement of the integration between virtual and real contents of the museum, in order to make them synergic and harmonious with the space, the paths, the visitors' needs [and] hybridization of media, bringing together different paradigms and languages of "representation" (virtual reality, theatre, cinema, sound and light design, mixed and augmented reality, holography, game), in order to involve and guide the visitor through a vivid and incisive narrative cultural experience'. These technologies may prospectively grant proxy access to many thousands of historic and cultural artefacts, including relics, that might otherwise be inaccessible. This, combined with the availability of public, occupational, and personal computational devices holds significant promise for increasing access to relics for public engagement and scholarly inquiry. As digital technologies continue to reshape the (museum) landscape, they introduce new layers of abstraction that can transform how relics are experienced and understood. While this can enhance accessibility and engagement, it also challenges traditional notions of authenticity and sensory experience, sensitivities around which have existed as long as relics have been around. These sensitivities are integral to the meaning and significance of relics.

## *5. Conclusion*

The nominal targeting of 'relics' has provided opportunities to consider properties of significant objects including their values, meanings, histories, and provenance, that might be captured and represented digitally. This leads us to present a set of lessons learned for capturing relics though hybrid technologies, which hold relevance for many other significant artefacts: (i) digital renderings of relics must acknowledge both their historical origins and contemporary significance in a way that (sensitively, i.e., according to the interaction context) balances perspectives; (ii) digital relics must be presented with clear information about their origin, 'authenticity', and significance, the sources of which must be transparent to the user; (iii) users must be enabled to engage with digital relics autonomously to respect and encourage agency and foster trust; (iv) commodification of digital relics in a way that exploits cultural or personal belief should be avoided; and (v) space for multiple perspectives must be provided in order to honour the diversity of values and prevent exclusionary biases. With the advent of AI and other data-driven technologies reliant on ground-truths, knowledge representation, and digital models of the real-world, these issues are especially salient as technologists grapple with post-truth politics, decolonisation, global conflict, responsible innovation, and identity politics, all of which require understandings of the social world from multiple-perspectives. Engagement across disciplinary boundaries is essential if we are to foster responsible innovation at the intersection of physical and digital spaces. Our research did not—indeed could not—seek to compile a definitive list of principles for incorporating faith in hybrid technologies. Rather, we consider them as catalysts for opening a dialogue from which a transdisciplinary understanding of these issues may develop. These may be interpreted and addressed differently from other disciplinary perspectives and as such we encourage wider engagement to build on this work.

## Statements and Declarations

### Acknowledgements

As we start to think about the entanglement of culture/language, technology and embodiment it becomes evident that we need to appreciate the interplay of data, the ways that this is embodied in the world and how we need to think about this in responsible ways using different critical framings.

This work was supported by the UKRI, EPSRC Trusted Data Driven Products [grant number EP/T022493/1), the Turing AI World Leading Researcher Fellowship in Somabotics: Creatively Embodying Artificial Intelligence [grant number EP/Z534808/1], AI UK: Creating an International Ecosystem for Responsible AI Research and Innovation (RAKE) [EP/Y009800/1] and the AHRC, Revision of the Anglo-Norman Dictionary (Letter V and Consolidation) [grant number AH/Z50743X/1].

**Competing Interests**
The authors have no competing interests to declare that are relevant to the content of this article.